\documentclass[3p,times,procedia]{elsarticle}
\usepackage{nupha_ecrc}

\volume{00}

\firstpage{1}

\journalname{Nuclear Physics A}

\runauth{}

\jid{nupha}

\jnltitlelogo{Nuclear Physics A}

\usepackage{amssymb}

\usepackage[figuresright]{rotating}

\begin{document}

\begin{frontmatter}




\dochead{XXVIth International Conference on Ultrarelativistic Nucleus-Nucleus Collisions\\ (Quark Matter 2017)}

\title{The right choice of moment for anisotropic fluid dynamics}


\author{H.\ Niemi}

\author{E.\ Moln\'ar}

\author{D.\ H.\ Rischke}
\address{Institut f\"ur Theoretische Physik, Johann Wolfgang Goethe-Universit\"at,
Max-von-Laue-Str.\ 1, D-60438 Frankfurt am Main, Germany}


\begin{abstract}
We study anisotropic fluid dynamics derived from the Boltzmann equation based 
on a particular choice for the anisotropic distribution function within a 
boost-invariant expansion of the fluid in one spatial dimension.
In order to close the conservation equations we need to choose an additional moment 
of the Boltzmann equation. We discuss the influence of this choice of closure 
on the time evolution of fluid-dynamical variables and search for the best agreement 
to the solution of the Boltzmann equation in the relaxation-time approximation.
\end{abstract}




\end{frontmatter}


\section{Introduction}
\label{sec:introduction}

The basic axioms of fluid dynamics are the conservation laws of particle number and energy-momentum, 
\begin{equation} \label{eom}
\partial_\mu N^\mu =0, \;\;\;\; \partial_\mu T^{\mu \nu} = 0 ,
\end{equation}
where $N^\mu$ is the particle four-current and $T^{\mu \nu}$ is the energy-momentum tensor.
These are decomposed with respect to the normalized fluid four-velocity $u^{\mu}$ and 
the projection orthogonal to it $\Delta^{\mu\nu} = g^{\mu\nu}-u^{\mu}u^{\nu}$, 
where $g^{\mu\nu}=\textrm{diag}(1,-1,-1,-1)$ is the metric of space-time, as 
\begin{equation}
\label{eq:N_mu}
N^{\mu} = n_0 u^{\mu} + V^{\mu}, \;\;\;\;
T^{\mu \nu} = e_0 u^{\mu} u^{\nu} - (P_0 + \Pi)\Delta^{\mu \nu}  + 2 W^{(\mu}u^{\nu)} + \pi^{\mu \nu},
\end{equation}
where $2 a^{(\mu} b^{\nu)} \equiv a^\mu b^\nu + a^\nu b^\mu$.
From these decompositions one can identify the following 14 quantities:
The particle density $n_0=N^{\mu} u_{\mu}$, energy density $e_0=T^{\mu \nu} u_{\mu} u_{\nu}$, 
and total pressure $P_0 + \Pi = -\frac{1}{3} T^{\mu \nu} \Delta_{\mu\nu}$, being the sum of 
$P_0(n_0,e_0)$ and the bulk viscous pressure $\Pi$. 
Furthermore, the particle diffusion current $V^{\mu}=\Delta^{\mu}_{\alpha} N^{\alpha}$, 
the energy diffusion current $W^{\mu} = \Delta _{\alpha }^{\mu }T^{\alpha \beta}u_{\beta }$, 
and the shear-stress tensor 
$\pi^{\mu\nu} = \Delta_{\alpha \beta }^{\mu \nu}T^{\alpha \beta }$, where 
$\Delta _{\alpha \beta }^{\mu \nu } =\frac{1}{2}\left( \Delta _{\alpha}^{\mu }\Delta _{\beta }^{\nu }
+\Delta _{\beta }^{\mu }\Delta _{\alpha}^{\nu }\right) -\frac{1}{3}\Delta ^{\mu \nu }\Delta _{\alpha \beta }$.

A sufficiently simple microscopic theory, where different assumptions for the derivation of the fluid-dynamical 
equations can be explicitly carried out and tested, is provided by the Boltzmann equation \cite{deGroot,Cercignani_book}
\begin{equation}\label{BTE}
 k^{\mu}\partial_{\mu} f_{\mathbf{k}} = C\left[f_{\mathbf{k}}\right],
\end{equation} 
where $f_{\mathbf{k}} = f(x,k)$ is the single-particle distribution function, $k^{\mu}$ is the four-momentum 
of particles, and $C\left[f_{\mathbf{k}}\right]$ is the collision integral. A conventional way to derive fluid dynamics is
to assume that the system is sufficiently close to local thermal equilibrium, characterized by a single-particle
distribution function
$f_{0\mathbf{k}}(\alpha_0, \beta_0)$, so that one can write $f_{\mathbf{k}} = f_{0\mathbf{k}} + \delta f_{\mathbf{k}}$, 
with $|\delta f_{\mathbf{k}}| \ll f_{\mathbf{k}}$, and then express $\delta f_{\mathbf{k}}$ in terms of only a few additional 
macroscopic quantities, e.g., as in Eq.\ (\ref{eq:N_mu}).

The inverse temperature $\beta_0=1/T$ 
and chemical potential over temperature $\alpha_0 =\mu/T$, which appear as parameters in $f_{0\mathbf{k}}$, 
are usually defined through the so-called Landau matching conditions, i.e., by requiring that
the particle density $n \equiv N^\mu u_\mu$ and energy density $e \equiv T^{\mu \nu} u_\mu u_\nu$ in the frame
where the 4-velocity $u^\mu= (1,0,0,0) $ are identical to those of the local equilibrium state,
$n = n_0(\alpha_0, \beta_0)$ and $e = e_0(\alpha_0, \beta_0)$. The Boltzmann equation can then be used to 
write down the equations of motion for the dissipative quantities appearing
in Eq.~(\ref{eq:N_mu}). For details of the expansion and the derivation 
see e.g., Refs.\ \cite{deGroot,Cercignani_book,Denicol:2012cn}.

Anisotropic fluid dynamics is based on a similar idea, but now the expansion is performed around a more 
general anisotropic distribution function
$\hat{f}_{0\mathbf{k}}(\alpha_0, \beta_0, \xi)$, as $f_{\mathbf{k}} = \hat{f}_{0\mathbf{k}} + \delta\hat{f}_{\mathbf{k}}$, 
instead of an isotropic equilibrium state.
Part of the possible deviations from the equilibrium distribution function $f_{0\mathbf{k}}$ 
can then be embedded into $\hat{f}_{0\mathbf{k}}$ 
\cite{Florkowski:2008ag,Martinez:2009ry,Florkowski:2010cf, Bazow:2013ifa, Bazow:2015zca}.
If the momentum anisotropy is large, this can lead to a much faster convergence of the expansion.
The degree of anisotropy in $\hat{f}_{0\mathbf{k}}$ is controlled by the new parameter $\xi$, 
while the direction of the anisotropy can be specified by a new spacelike 4-vector $l^{\mu}$ orthogonal 
to the four-velocity, $u^{\mu}l_{\mu}=0$, and normalized to $l^{\mu}l_{\mu} = -1$. 
The additional parameter $\xi$ needs to be determined by another matching condition in addition to 
the usual Landau matching conditions.

Now, $N^{\mu}$ and $T^{\mu\nu}$ need to be decomposed with respect to both four-vectors $u^{\mu}$ 
and $l^{\mu}$ as well as the 
two-space projector $\Xi^{\mu \nu } \equiv g^{\mu \nu }-u^{\mu }u^{\nu}+l^{\mu }l^{\nu }$ orthogonal to 
both four-vectors \cite{Barz:1987pq}. This reads as
\begin{equation}
\nonumber
N^{\mu } =n\, u^{\mu }+n_{l}\, l^{\mu }+V_{\perp }^{\mu },
\label{kinetic:N_mu_u_l} 
\end{equation}
\begin{equation}
\nonumber
T^{\mu \nu } = e\, u^{\mu }u^{\nu }+2\, M\, u^{\left( \mu \right. }l^{\left. \nu\right) }
+P_{l}\, l^{\mu }l^{\nu }-P_{\perp }\, \Xi ^{\mu \nu }+2\, W_{\perp u}^{\left( \mu \right. }
u^{\left. \nu \right) }+2\, W_{\perp l}^{\left( \mu\right. }l^{\left. \nu \right) }
+\pi _{\perp }^{\mu \nu },
\end{equation}
where compared to Eq.\ (\ref{eq:N_mu}) the particle diffusion current $V^\mu$ splits into two parts: 
The diffusion into the direction of $l^{\mu}$ given by $n_l=-N^{\mu}l_{\mu}$ and
the part of the diffusion orthogonal to $l^{\mu}$ given by $V_{\perp }^{\mu}=\Xi^{\mu}_{\alpha} N^{\alpha}$. 
Similarly, the isotropic pressure splits into longitudinal and transverse
parts $P=\frac{1}{3}\left( P_{l}+ 2P_{\perp }\right)$, where 
$P_{l} =T^{\mu \nu}l_{\mu }l_{\nu }$ and $P_{\perp } = -\frac{1}{2}\, T^{\mu \nu }\Xi _{\mu \nu }$, 
and thus the shear-stress tensor from the second equation (\ref{eq:N_mu}) can be written as
\begin{equation}
 \pi ^{\mu \nu }=\pi _{\perp }^{\mu \nu }+2\, W_{\perp l}^{\left( \mu \right.}l^{\left. \nu \right) }
+\frac{1}{3}\left( P_{l}-P_{\perp }\right) \left(2\, l^{\mu }l^{\nu }+\Xi ^{\mu \nu }\right),  
\label{pi=pi_t+W_l+P_tP_l}
\end{equation}
where $W_{\perp l}^{\mu } =-\Xi _{\alpha }^{\mu }T^{\alpha \beta }l_{\beta }$ and 
$\pi _{\perp }^{\mu \nu } =\Xi _{\alpha \beta }^{\mu \nu}T^{\alpha \beta }$, with
$\Xi _{\alpha \beta }^{\mu \nu } =\frac{1}{2}\left( \Xi _{\alpha}^{\mu }\Xi _{\beta }^{\nu }
+\Xi _{\beta }^{\mu }\Xi _{\alpha}^{\nu }\right) -\frac{1}{2}\Xi ^{\mu \nu }\Xi _{\alpha \beta }$.
Similarly to conventional fluid dynamics, the equations of motion for the anisotropic dissipative quantities can be also obtained from the Boltzmann equation, see e.g.\ Ref.\ \cite{Molnar:2016vvu}.

\section{Boost-invariant expansion}
\label{sec:boost_invariant}

\begin{figure}[]
\includegraphics[width=0.5\textwidth]{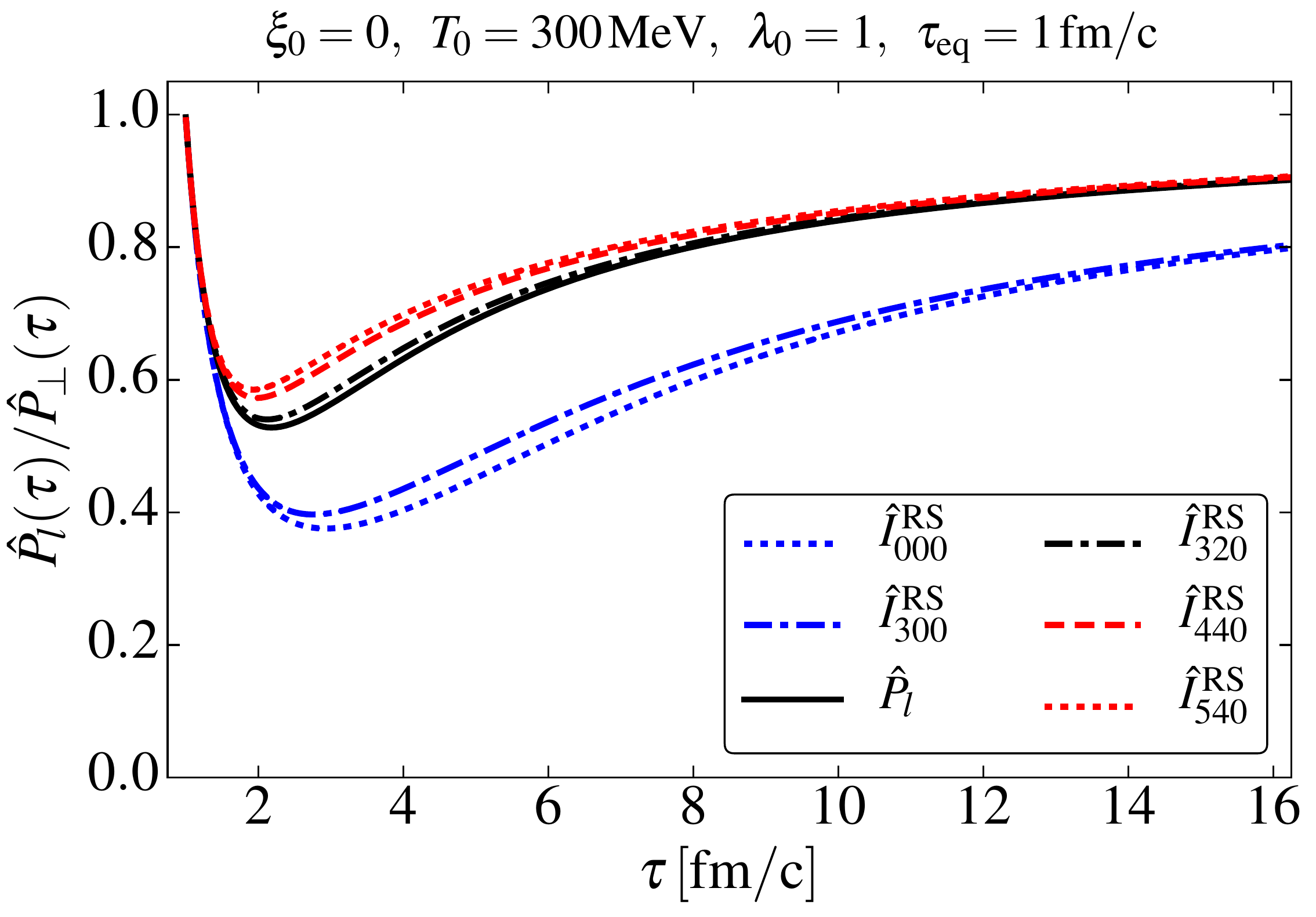} 
\includegraphics[width=0.5\textwidth]{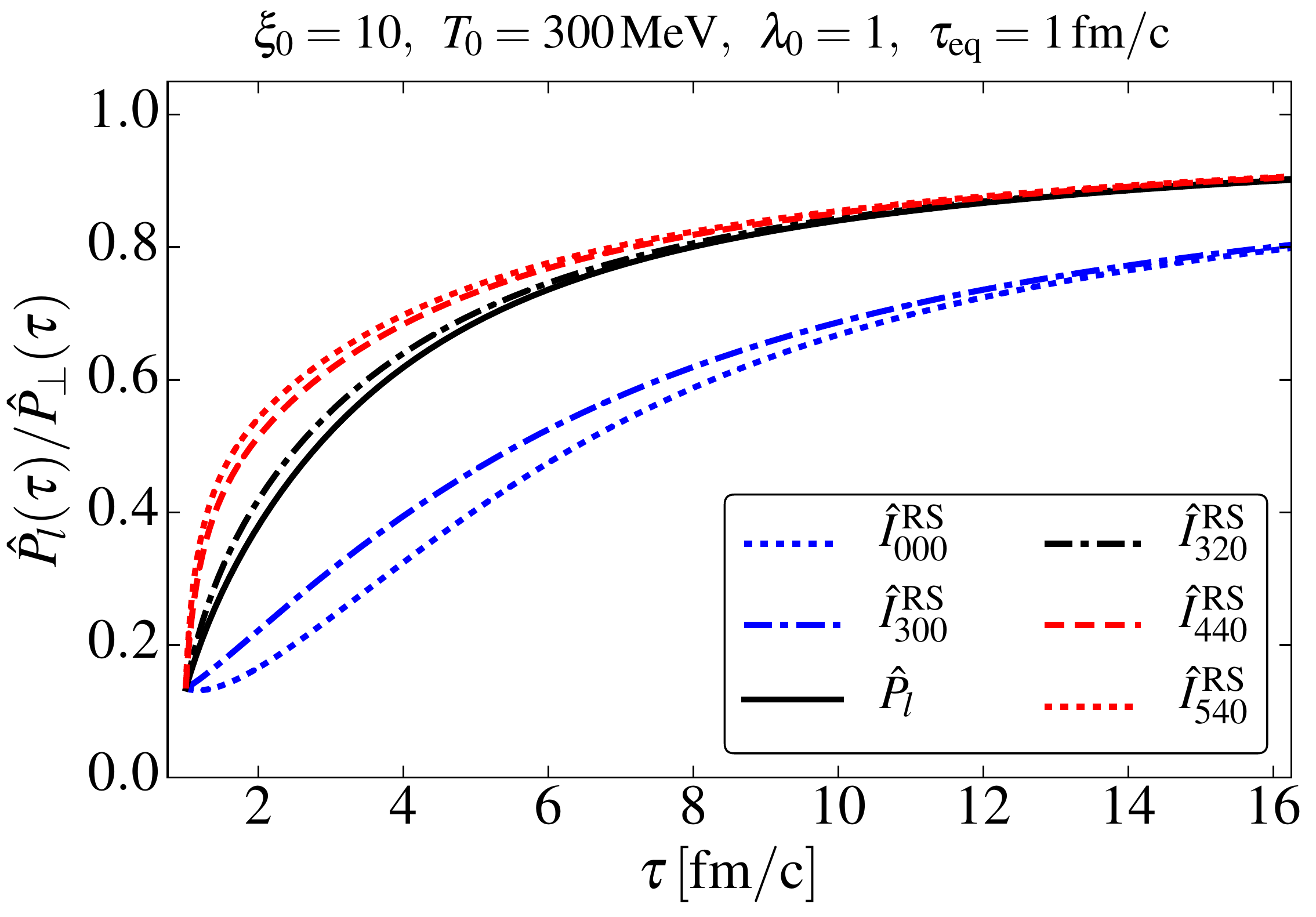} 
\vspace{-0.5cm}
\caption{The evolution of $\hat{P}_l/\hat{P}_{\perp}$ as a function of proper time $\tau$ for the different choices of $\hat{I}^{RS}_{nrq}$ for closure. The figure is taken from Ref.~\cite{Molnar:2016gwq}.}
\label{fig:choices}
\end{figure}

As a simple example to illustrate the importance of choosing the right matching, we utilize the anisotropic
distribution function introduced by Romatschke and Strickland (RS) \cite{Romatschke:2003ms},
\begin{equation} \label{eq:B01_hatf0_anistrop}
\hat{f}_{0\mathbf{k}} \equiv \hat{f}_{RS} =  \left[ \exp \left( \beta_0\sqrt{E_{\mathbf{k} u}^2 
+ \xi E_{\mathbf{k}l}^2} - \alpha_0 \right)+ a \right]^{-1},
\end{equation}
where $E_{\mathbf{k}u} \equiv u^\mu k_\mu$ and $E_{\mathbf{k}l} \equiv - l^\mu k_\mu$. Note that when $\xi = 0$, this
reduces to an equilibrium distribution function.
We then solve anisotropic fluid dynamics in a simple 0+1 dimensional boost-invariant geometry, and 
compare the solution to the exact solution of the Boltzmann equation in the relaxation-time 
approximation (RTA)~\cite{Molnar:2016gwq}.
A general moment of the distribution function (\ref{eq:B01_hatf0_anistrop}) can be written as
\begin{equation}
\hat{I}_{nrq}^{RS}=\frac{\left( -1\right) ^{q}}{\left( 2q\right) !!}\int
dK\, E_{\mathbf{k}u}^{n-r-2q}E_{\mathbf{k}l}^{r}\left( \Xi ^{\mu \nu }k_{\mu}k_{\nu }\right)^{q}\ \hat{f}_{RS}\;,  \label{I_nrq_RS}
\end{equation}
hence for example $\hat{n} = \hat{I}_{100}^{RS}$, $\hat{e} = \hat{I}_{200}^{RS}$, and 
$\hat{P}_{l} = \hat{I}_{220}^{RS}$, etc. 
Now, imposing the Landau matching conditions $\hat{n}(\alpha_{RS},\beta_{RS},\xi)=n_0(\alpha_0,\beta_0)$
and  $\hat{e}(\alpha_{RS},\beta_{RS},\xi)=e_0(\alpha_0,\beta_0)$ we determine two parameters of the 
anisotropic distribution in terms of a fictitious equilibrium state.

In the 0+1 dimensional boost-invariant expansion the conservation laws take the simple form
\begin{equation}
\frac{\partial n_{0}\left( \alpha _{0},\beta _{0}\right) }{\partial \tau }+
\frac{1}{\tau }n_{0}\left( \alpha _{0},\beta _{0}\right) =0 , \;\;\;\;
\frac{\partial e_{0}\left( \alpha _{0},\beta _{0}\right) }{\partial \tau }
+\frac{1}{\tau }\left[ e_{0}\left( \alpha _{0},\beta _{0}\right) +\hat{P}_{l}
\left( \alpha _{RS},\beta _{RS},\xi \right) \right] =0 .
\label{BJ_e_cons}
\end{equation}
The equations of motion can be closed by providing an additional equation for one of the moments $\hat{I}_{nrq}^{RS}$,
which, with the help of the Boltzmann equation in RTA, can be written as
\begin{equation}
\frac{\partial \hat{I}^{RS}_{i+j,j,0}}{\partial \tau }+\frac{1}{\tau }\left[
\left( j+1\right) \hat{I}^{RS}_{i+j,j,0}+\left( i-1\right) \hat{I}^{RS}_{i+j,j+2,0}
\right] = -\frac{1}{\tau _{eq}} 
\left( \hat{I}^{RS}_{i+j,j,0} - I_{i+j,j,0}\right) ,
\label{Main_eq_motion}
\end{equation}
where $I_{i+j,j,0} =\int dK\, E_{\mathbf{k}u}^{i+j}E_{\mathbf{k}l}^{j}\ f_{0\mathbf{k}}$.  
For the relaxation time $\tau_{eq}$ we will either use a constant value, 
or parametrize it using the relation between $\tau_{eq}$ and shear viscosity \cite{Florkowski:2013lza,Florkowski:2013lya},
$\tau_{eq}(\tau) = 5\beta_0(\tau)\eta/s$,
where $\eta/s$ denotes a constant ratio of shear viscosity to entropy density. As an example we can write the 
equation of motion directly for the longitudinal pressure that appears in the conservation 
laws (\ref{BJ_e_cons}), 
\begin{equation}
\frac{\partial \hat{P}_{l}}{\partial \tau }+\frac{1}{\tau }\left( 3\hat{P}_{l}-\hat{I}_{240}^{RS}\right) 
=-\frac{1}{\tau _{eq}}\left( \hat{P}_{l}-P_{0}\right)  ,  \label{BJ_Pl_relax}
\end{equation}
or for the moment $\hat{I}_{300}^{RS}$,
\begin{equation}
\frac{\partial \hat{I}_{300}^{RS}}{\partial \tau }+\frac{1}{\tau }\left( 
\hat{I}_{300}^{RS}+2\hat{I}_{320}^{RS}\right) 
=-\frac{1}{\tau _{eq}}\left( \hat{I}_{300}^{RS}-I_{300}\right),  \label{BJ_I300_relax}
\end{equation}
that also closes the system. All moments $\hat{I}_{nrq}^{RS}$ are related through the chosen 
anisotropic distribution function, i.e.,
once $\alpha_0$, $\beta_0$, and $\xi$ are given, all these moments are determined. 
Further choices are shown in Fig.\ \ref{fig:choices}.
By choosing one of the moment equations to close the system also implies a matching condition that 
determines the parameter $\xi$.  
If the system is closed using Eq.~(\ref{BJ_Pl_relax}), it implies that $\alpha_0$, $\beta_0$, and $\xi$ are matched 
to $\hat{P}_l \equiv \hat{I}^{RS}_{220}$, 
and choosing Eq.~(\ref{BJ_I300_relax}) implies that they are matched to $\hat{I}^{RS}_{300}$. 
Note that in general different choices 
lead to different values for $\alpha_0$, $\beta_0$, and $\xi$. The longitudinal pressure appears directly in the 
conservation laws, therefore it
is a natural choice for matching the anisotropy parameter. As it turns out, this choice gives also the best 
agreement with the exact solution of the Boltzmann equation.

In Fig.~\ref{fig:choices} we show the evolution of the ratio of longitudinal and transverse pressure components 
$\hat{P}_l/\hat{P}_{\perp}$  
as a function of proper time $\tau$ for different choices of $\hat{I}^{RS}_{nrq}$ to close 
the system~\cite{Molnar:2016gwq}. The relaxation time is taken
as a constant $\tau_{eq} = 1$ fm, and the initial anisotropy parameter is $\xi_0 = 0$ or $10$. 
As can be seen from the figure, the different choices lead to quite large differences in the evolution. 
Thus, the right choice of moment becomes essential in order
to reliably describe the evolution of the system.

\begin{figure}[]
\includegraphics[width=0.5\textwidth]{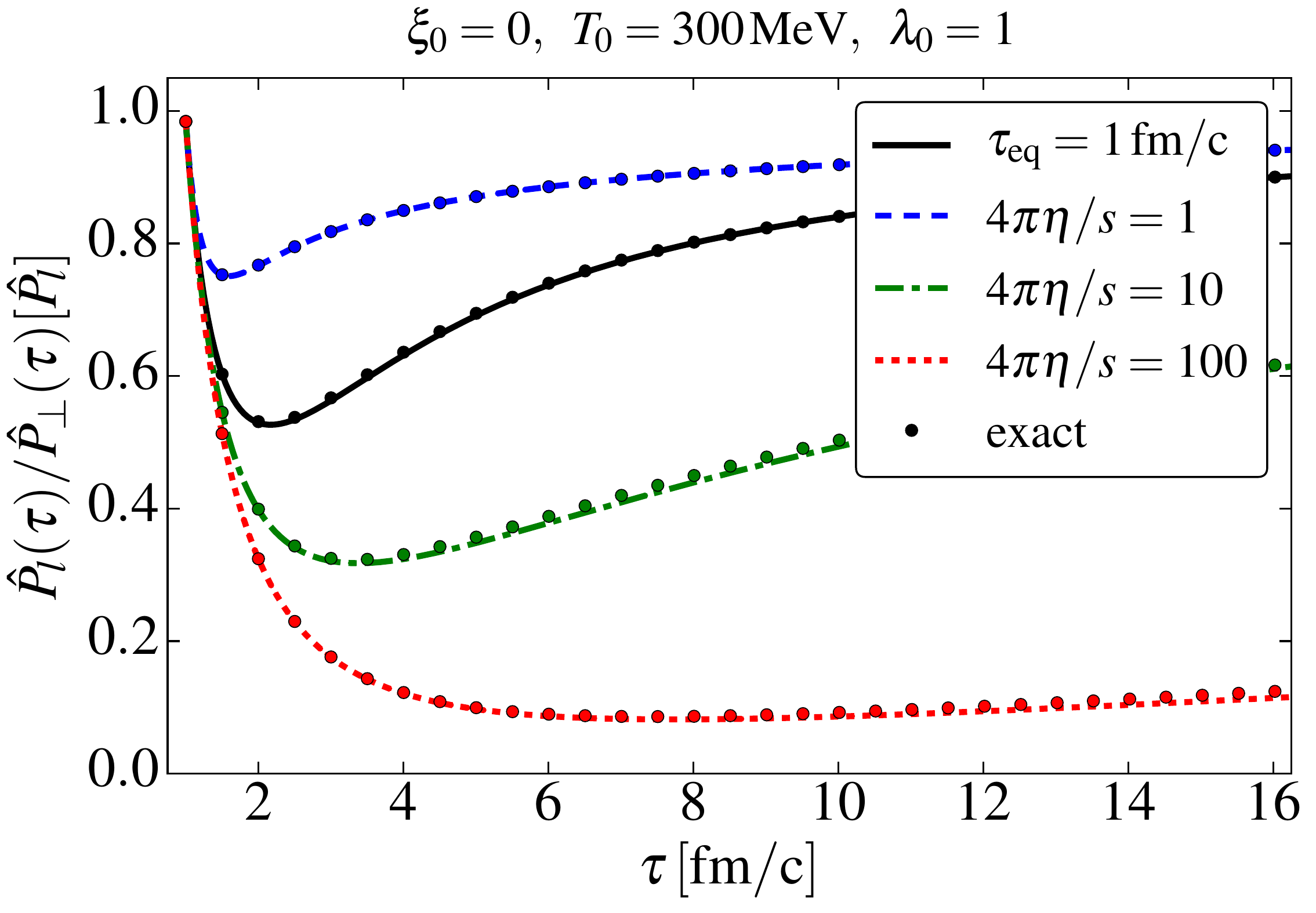} %
\includegraphics[width=0.5\textwidth]{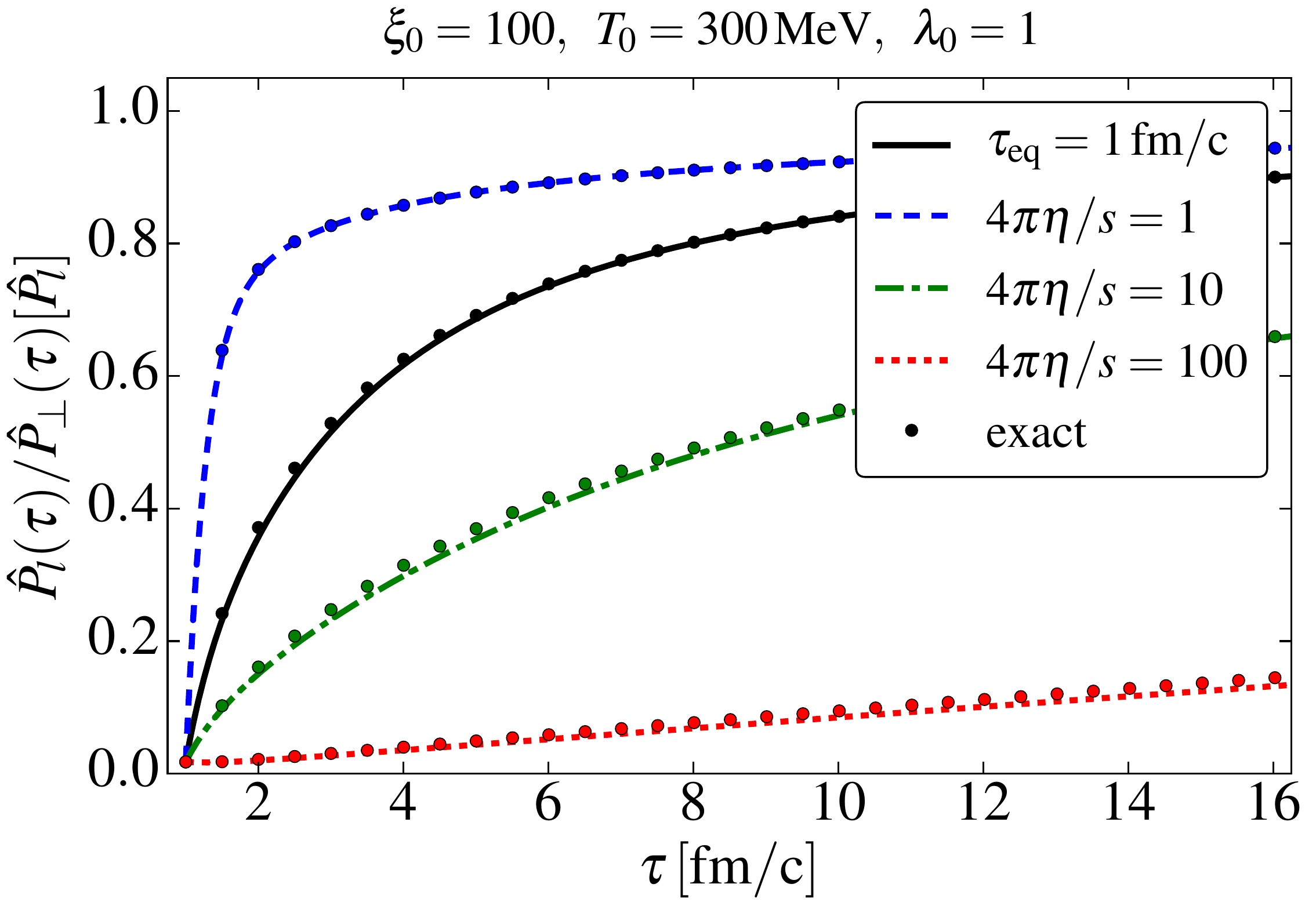} %
\vspace{-0.5cm}
\caption{(Color online) The evolution of $\hat{P}_l/\hat{P}_{\perp}$ as a function of proper time $\tau$.
The lines are the solution of the conservation equations closed by the relaxation equation for 
$\hat{P}_l$ with $\tau_{eq}=1$ fm and with three different choices for $\eta/s$.
The corresponding solutions of the Boltzmann equation are shown by the large dots. The figure is taken
from Ref.~\cite{Molnar:2016gwq}.}
\label{fig:exact}
\end{figure}

The best choice for matching can be found by comparing to the solution of the Boltzmann equation which 
can be solved exactly in RTA~\cite{Florkowski:2013lza,Florkowski:2013lya}. In Fig.~\ref{fig:exact}
we show the solution of anisotropic fluid dynamics with Eq.~(\ref{BJ_Pl_relax}) as choice for closing the system. 
This is the choice that corresponds to the one of Ref.~\cite{Tinti:2015xwa}, and in conventional fluid dynamics to the 
one of Ref.~\cite{Denicol:2010xn}.
The evolution is shown for several 
values of $\tau_{eq}$ or $\eta/s$, and for two different values of the initial anisotropy parameter $\xi_0$. 
The agreement between the 
two approaches is excellent and persists up to very large $\eta/s$ and large values of the initial anisotropy. This shows
that matching to the longitudinal pressure gives the overall best agreement with the Boltzmann 
equation~\cite{Molnar:2016gwq}. The other choices
shown in Fig.~\ref{fig:choices} can lead to strong deviations from the Boltzmann equation.

\section{Acknowledgments}

The work of E.M.\ was  supported by BMBF grant no.\ 05P15RFCA1. 
H.N.\ has received funding from the European Union's Horizon 2020 research and innovation
programme under the Marie Sklodowska-Curie grant agreement no.\ 655285.
E.M.\ and H.N.\ were partially supported by the Helmholtz International Center for
FAIR within the framework of the LOEWE program launched by the State
of Hesse.

\end{document}